# Polarization Enhancement in Perovskite Superlattices by Oxygen Octahedral Tilts

X. Z. Lu[a)], X. G. Gong, and H. J. Xiang[b)]

Key Laboratory of Computational Physical Sciences (Ministry of Education), State Key Laboratory of Surface Physics, and Department of Physics, Fudan University, Shanghai 200433, P. R. China

e-mail: a) xuezenglu11@fudan.edu.cn; b) hxiang@fudan.edu.cn

**Abstract**

Interface engineering in perovskite oxide superlattices has developed into a flourishing field, enabling not only further tuning of the exceptional properties, but also giving access to emergent physical phenomena. Here, we reveal a new mechanism for enhancing the electric polarization by the interface-induced oxygen octahedral tilts in $BaTiO_3$/$CaTiO_3$ superlattices. By combining a novel genetic algorithm with density functional theory (DFT), we predict that the true ground states in 1:1 and 2:2 $BaTiO_3$/$CaTiO_3$ superlattices grown on $SrTiO_3$ adopt *Pc* symmetry with a large proper electric polarization (32.8μC/cm$^2$ for 1:1 and 35.8 μC/cm$^2$ for 2:2 superlattices), which is even larger than that of bulk $BaTiO_3$. The tilt of oxygen octahedron is found to play a key role for the enhancement of out-of-plane polarization in 1:1 superlattices because it reduces greatly the rotation of oxygen octahedron (out-of-phase) which significantly suppresses the out-of-plane polarization.

# 1. Introduction

Interfaces formed by perovskite oxides offer a tremendous opportunity for fundamental as well as applied research. In this active field of research, interfacial effects on the polarization of the ultra-short superlattices have attracted enormous interests [1-8]. Considering only the out-of-plane ferroelectric modes, Neaton *et al*. predicted that the polarization of BaTiO$_3$/SrTiO$_3$ superlattices (*P4mm* structure) with BaTiO$_3$ (BTO) fraction larger than 40% grown on SrTiO$_3$ (STO) substrate was enhanced above that of bulk BTO [1]. The effects of oxygen octahedral rotations around [001] [an antiferrodistortive (AFD) mode] on the ferroelectricity were examined later: It was shown that a trilinear coupling between the ferroelectric and interfacial (in-phase and out-of-phase) AFD rotation modes results in an enhanced polarization in the 1:1 PbTiO$_3$/SrTiO$_3$ (PTO/STO) superlattices with octahedral rotations (*P4bm* structure) when compared to *P4mm* structure [7,9]. Recently, the role of another common AFD mode, i.e., oxygen octahedral tilt (rotation of oxygen octahedron around an axis perpendicular to [001]) played on the polarization has been elucidated [10-15], in which the in-plane polarization was induced by another type of trilinear coupling between the polarization, oxygen octahedral rotation and tilt. The polarization resulting from the trilinear coupling was then referred as hybrid improper ferroelectricity (HIF) [10]. According to this model, Mulder *et al*. [13] raised a strategy to design artificial oxides with large electric polarization (*P*) and small energetic switching barriers between +*P* and −*P*.

In this work, we investigate a typical ferroelectric/dielectric superlattice system,

i.e., $(BaTiO_3)_n/(CaTiO_3)_n$ (n is the number of layers). Previous first principles studies showed that the polarization of $(BaTiO_3)_2/(CaTiO_3)_2$ on STO substrate with the *P4mm* structure was larger than that of the tetragonal bulk BTO [16]. But it decreased greatly in *P4bm* structure and became even smaller than the tetragonal bulk BTO value. This decreasing was attributed to the suppressing of the polarization by an oxygen octahedral rotation [17,18]. It should be noted that all previous studies neglected the tilts of oxygen octahedra, which is one of the most important distortions in bulk *Pbnm* CaTiO$_3$.

In order to consider all possible relevant distortions in perovskites, we propose a global optimization method based on the genetic algorithm (GA) (see Computational Methods) to search the lowest energy structures of BTO/CTO superlattices. By combining the GA with density functional theory (DFT) calculations, the true ground states of 1:1 and 2:2 BTO/CTO superlattices are predicted to adopt the *Pc* symmetry. We find that the enhanced out-of-plane polarization in 1:1 *Pc* structure with the oxygen octahedral tilt occurs because the tilt of oxygen octahedra suppresses dramatically the oxygen octahedral rotation (out-of-phase), which disfavors the out-of-plane ferroelectricity. Finally due to the existence of a large in-plane polarization, the total polarizations of 1:1 and 2:2 BTO/CTO on STO are predicted to be 32.8 μC/cm$^2$ and 35.8 μC/cm$^2$ respectively, which are larger than the bulk value of tetragonal BTO.

2. **Computational methods**

Since the 1990s, genetic algorithms (GAs) have been used to search the ground state of nanoclusters [19,20], alloys [21,22], and crystals [23]. They employ a search technique based on principles similar to those of natural selection, singling out the most "adaptive" structures, which have the lowest energies [19,20,23]. In this study, we are dealing with a problem different from the previous studies: Here, the basic lattice structure (i.e., perovskite based structures) is fixed and our purpose is to find out the distortion of the cell and ionic positions which leads to the lowest energy. To this end, we propose a genetic algorithm which differs from the previous algorithms in several aspects: (1) To generate an initial structure of the first generation, we first randomly select a subgroup of the space group of the undistorted perovskite structure for a given superstructure. By symmetrizing a structure with random distortions using the symmetry operation of the subgroup, we can obtain an initial structure with this selected subgroup symmetry. (2) For the mating operation, we propose another crossover operation besides the usual cut-and-splice method proposed by Deaven and Ho [19]: In a cubic perovskite system, there are different unstable phonon modes such as the ferroelectric displacement and the oxygen octahedral rotation. The distortion in the ground state is usually a superposition of different modes. Guided by this physical insight, we introduce the following mating operation: $X_{child} = X_{cubic} + (X_{father} - X_{cubic}) + c(X_{mother} - X_{cubic})$, where $X_{father}$ and $X_{mother}$ are the two parent structures, $X_{cubic}$ and $X_{child}$ are the undistorted cubic structure and new child structure, respectively. For the new distortion, we take either the sum or difference of these two distortions depending on a constant c, which is randomly chosen to be 1 or -1.

Our total energy calculations are based on the density functional theory (DFT) [24] within the local density approximation (LDA) on the basis of the projector augmented wave method [25] encoded in the Vienna ab initio simulation package [26]. Calcium 3s, 3p and 4s electrons, Ba 5s, 5p and 6s electrons, Ti 3p, 3d and 4s electrons and O 2s and 2p electrons are treated as valence states. The plane-wave cutoff energy is set to 600 eV for calculating phase diagram and 450 eV for searching the ground state using the GA method. And $4 \times 4 \times 3$ $k$-point mesh is used for the 20-atom $\sqrt{2} \times \sqrt{2} \times 2$ cell and $4 \times 4 \times 1$ $k$-point mesh for the 40-atom $\sqrt{2} \times \sqrt{2} \times 4$ cell. For the electric polarization calculations, the berry phase method [27] is used.

**3. Results and discussion**

The lattice vectors used in this study are $\mathbf{a} = a_s\mathbf{x} - a_s\mathbf{y}$, $\mathbf{b} = a_s\mathbf{x} + a_s\mathbf{y}$ and $\mathbf{c} = \delta_1\mathbf{x} + \delta_2\mathbf{y} + (2a_s + \delta_3)\mathbf{z}$ for 20-atom cell and $\mathbf{c} = \delta_1\mathbf{x} + \delta_2\mathbf{y} + (4a_s + \delta_3)\mathbf{z}$ for 40-atom cell, where $a_s$ is the in-plane lattice constant of the studied system at a given strain, and (**x**, **y**, **z**) are defined within the pseudocubic setting. Epitaxial strain is then defined as $(a_s - a_0)/a_0$, where $a_0$ (3.856 Å) is the theoretical lattice constant of bulk STO within LDA optimization in Ref. [17].

The total energies of the lowest energy phases in 1:1 BTO/CTO superlattices from the GA simulations, as a function of epitaxial strain from −4% to 4%, are shown in Figure 1a. For comparison, the previous suggested *P4bm*, *P4mm* and *Pm* [28] phases are also shown. Fig. 1b displays the dependence of the total polarizations of the lowest energy phases on the epitaxial strain for 1:1 superlattices. The results for 2:2 superlattices are given in Ref. [29].

As can be seen in Fig. 1a, in the strain range of -4% - 1%, *Pc* phase is stabilized, which is lower in energy by an average of 14 meV per formula unit with respect to *P4bm* phase. This is also confirmed by the phonon calculations indicating that *P4bm* structure is unstable. In *Pc* structure, the tilts of oxygen octahedra and the in-plane ferroelectric distortion occur which lower the energy when compared with *P4bm* structure. These rotations and tilts of oxygen octahedra related to CTO bulk-like layers can be attributed to the interfacial effects, which do not exist in BTO bulk. As the strain is higher than 1%, *Pc* phase is gradually transformed into *Pmc*$2_1$ phase. The phase diagram for 2:2 superlattices is similar to that of 1:1 superlattices, but the stable region is -4% - 2% for *Pc* phase and 2% - 4% for *Pmc*$2_1$ phase (see Fig. S1a of Ref. [29]). Although these two phases were also found in 1:1 and 2:2 PTO/STO superlattices [9], the range of *Pc* phase is much wider in BTO/CTO, i.e. -4% - 1% for 1:1 superlattices and -4% - 2% for 2:2 superlattices, to be compared with -1% – 1% for 2:2 PTO/STO and a negligible range for 1:1 PTO/STO [9]. This is due to the fact that the instabilities of the tilt of oxygen octahedron for 1:1 superlattices and the tilt of oxygen octahedron and in-plane ferroelectric distortion for 2:2 superlattices still exist at -4% strain in our case (see Table I and Table SI of Ref. [29]). The monoclinic *Pc* structure has both in-plane and out-of-plane ferroelectric components. As shown in Table II, the out-of-plane polarization in *Pc* structure is larger than that in *P4bm* structure at 0% strain and will gradually increase with increasing the compressive strain (see Fig. 1b). Combined with the in-plane component, the total polarization for BTO/CTO (32.8 μC/cm$^2$) when grown on [001] STO is even larger than that (25.4 μC/cm$^2$) of the tetragonal BTO bulk from our theoretical calculation.

To find out the origins of this enhancement of out-of-plane polarization in 1:1 BTO/CTO at 0% strain, we carefully investigate the distortions in *Pc* structure. The centrosymmetric *P*4*mmm* structure is chosen to be the reference structure. We find that $\Gamma^{5-}$, $\Gamma^{3-}$, $M^{5-}$, $M^{3+}$ and $M^{1-}$ modes are condensed in this *Pc* structure, which correspond to in-plane (FE$_{xy}$) and out-of-plane (FE$_z$) ferroelectric modes, tilt of oxygen octahedron (AFD$_{xy}$), in-phase (AFD$_{zi}$) and out-of-phase (AFD$_{zo}$) rotations of oxygen octahedra, respectively (see Fig. 2). Phonon calculations show that all these modes are unstable (see Table I). Interestingly, the magnitude of the phonon frequency for AFD$_{xy}$ mode (oxygen octahedral tilt) is the largest, indicating that it is more likely to be one of the main structural distortions which drive the phase transition.

It has been known that there might exist trilinear coupling in the Landau expansion of free energy in the presence of the oxygen octahedral rotation and tilt that can induce the ferroelectricity in the artificial superlattices of perovskite oxides [11-15]. To know whether it is also applicable to our system, we firstly study the origin of ferroelectricity in 1:1 *Pc* structure by calculating the phonon frequencies of *P*4*mmm* structure at some strains. From Table I, we can see that from -4% to 4% the frequency of FE$_z$ mode is changed from negative to positive value, while the frequency of FE$_{xy}$ mode is changed in the opposite direction. Combined with the evolution of the polarization with the strain in Fig. 1b, it can be confirmed that FE$_z$ polarization should be proper ferroelectricity. Its evolving can be attributed to the polarization-strain coupling. For $FE_{xy}$ polarization, we can see that even in the absence of polar instability at -4% strain, the system still has a large $FE_{xy}$ polarization with a value of 9.2μC/cm$^2$, which should be induced by the symmetry-allowed $AFD_{zi}$ and $AFD_{xy}$ modes through the trilinear coupling, that is, it is hybrid improper ferroelectricity. Furthermore, a slightly decreasing in $FE_{xy}$ polarization can be observed in the range of -4 - -2% in Fig 1b, which cannot be attributed to the polarization-strain coupling as $FE_z$ polarization. Above -2% strain, the polarization starts to increase. With the changing in the phonon frequency of $FE_{xy}$ mode shown in Table I, we can expect that it has a transformation from hybrid improper ferroelectricity to proper ferroelectricity in $FE_{xy}$ polarization between -4% and -2% strain. Therefore, in the large compressive strain region, there is a coexsitence between the proper ferroelectricity and hybrid improper ferroelectricity. When grown on STO substrate, the polarization of 1:1 *Pc* structure is proper ferroelectricity.

Previously, it was suggested [17] that the rotations of oxygen octahedra suppress the out-of-plane polarization ($FE_z$) in BTO/CTO superlattices. We next investigate the effects of $AFD_{zi}$ and $AFD_{zo}$ rotations on $FE_z$ in more details. $FE_z$ polarization as a function of $AFD_{zi}$ and $AFD_{zo}$ rotations are calculated in which the internal atoms are relaxed according to $FE_z$ mode with the fixed $AFD_{zi}$ and $AFD_{zo}$ rotations. As can be seen in Fig. 3, $FE_z$ polarization is mainly influenced by $AFD_{zo}$ rotations and decreases greatly with increased the rotations.

We then investigate the effects of $FE_{xy}$, $AFD_{xy}$ and $AFD_{zi}$ modes on $AFD_{zo}$ mode. We first consider the case where only $AFD_{zo}$ is added to the reference *P4mmm* structure. Then we also add $FE_{xy}$, $AFD_{xy}$ and $AFD_{zi}$ modes respectively with the same values as that in *Pc* structure. As shown in Fig. 4a, only $AFD_{xy}$ tilts influence $AFD_{zo}$ rotations greatly, while the others hardly have any effect. Through the same processes, we also study the effects of $FE_{xy}$ and $AFD_{xy}$ modes on $FE_z$ mode, which is shown in Fig. S3 of Ref. [29]. It can be seen that the condensed $FE_{xy}$ and $AFD_{xy}$ modes in *Pc* structure have little effects on $FE_z$ polarization. As discussed above, $AFD_{zo}$ rotations will be only suppressed by $AFD_{xy}$ tilts. As a consequence, $FE_z$ polarization would increase in the presence of $AFD_{xy}$ tilt because it suppresses $AFD_{zo}$ rotation greatly (see Fig. 4b). Therefore, $AFD_{xy}$ mode plays an important role in enhancing $FE_z$ polarization in *Pc* structure of 1:1 BTO/CTO.

Why is the coupling (see Part 3 of Ref. [29]) between $AFD_{xy}$ mode and $AFD_{zo}$ mode so strong? In order to understand the microscopic origin of this coupling, we add $AFD_{xy}$ and $AFD_{zo}$ modes into the reference structure and consider two cases (see Fig. 5): (i) $\theta_Z^O = 3.1°$ and $\theta_{XY} = 3.4°$; (ii): $\theta_Z^O = 6.2°$ and the same $\theta_{XY}$ as in case (i). Because there exists only rotations and tilts of oxygen octahedra, the lengths of Ti-O bonds will remain almost unchanged. Thus we only examine the differences of the corresponding Ba-O and Ca-O (A-O) bond lengths between these two cases. Because A-O interaction is manly ionic, we firstly investigate the effects of the electrostatic interaction in these two cases. We use Coulomb interaction model of point charges to consider the electrostatic energy, which is calculated by Ewald technique [31]. The results show that the electrostatic energy is -734.60 eV for case (i) and -733.75 eV for case (ii). Secondly, we also consider the effect of Pauli repulsion when A-O bond distance becomes shorter than the sum of the ionic radius (2.75 Å for Ba-O bond and 2.4 Å for Ca-O bond [32]). As shown in Fig. 5, we find the shorter Ba-O bonds lengths in case (ii) are all less than those in case (i), which indicates that the case (ii) will have stronger repulsion if it exists. Due to the ionic interaction between A-ion and oxygen atom, the system will decrease $AFD_{zo}$ rotations in the presence of $AFD_{xy}$ tilts in order to lower the electrostatic energy and reduce the repulsion. This mechanism is different from the covalent nature of the different $AFD_z$ rotations in the two interfaces for 2:2 PTO/STO systems in which there exists hybridization between the active lone pair of lead and oxygen atom [9].

Our proposed mechanism is different from the previous study of oxygen octahedral tilt induced ferroelectricity where the tilt influences the in-plane polarization (HIF) through the trilinear coupling [10-15]. In our case, we can expect to control the out-of-plane polarization (proper ferroelectricity) through the oxygen octahedral tilt suppressing the oxygen octahedral rotation. Moreover, this tilt may be used to tune the magnetic properties similar to the cases in layered perovskite $Ca_3Mn_2O_7$[10] and $BiFeO_3/LaFeO_3$ artificial superlattices [15]. Therefore this mechanism may provide a new route to realize the magnetoelectric couping.

## 4. Conclusions

In summary, a comprehensive study is carried out to investigate BTO/CTO superlattices. We first propose a global optimization method based on the genetic algorithm (GA) to search the ground state structures of superlattices. By using this method, the true ground states for 1:1 and 2:2 BTO/CTO superlattices are predicted to be *Pc* structures. Then, we find that FE$_z$ polarization in *Pc* structure is larger than that in *P*4*bm* structure. For 1:1 superlattices, this is due to the fact that the tilt of oxygen octahedra suppresses the rotation of oxygen octahedral (out-of-phase) that disfavors the out-of-plane polarization. Finally due to the existence of a large in-plane polarization, the total polarizations (proper ferroelectricity) are predicted to be 32.8 μC/cm$^2$ for 1:1 and 35.8 μC/cm$^2$ for 2:2 superlattices when grown on [001] STO substrate respectively, which are larger than the tetragonal bulk BTO value. Our work suggests that the oxygen octahedral tilt provides a new handle for tuning the exceptional properties in superlattices.

**Acknowledgements**

We acknowledge Prof. Xifan Wu and Mr. Hongwei Wang for useful discussions. Work at Fudan was partially supported by NSFC, the Special Funds for Major State Basic Research, Foundation for the Author of National Excellent Doctoral Dissertation of China, The Program for Professor of Special Appointment at Shanghai Institutions of Higher Learning, Research Program of Shanghai municipality and MOE.

ferroelectric distortions and oxygen octahedral rotations and tilts.

Table I. Calculated frequencies of modes [30] at Γ and M in *P4mmm* structure of 1:1 BTO/CTO superlattices at different strains.

| Strain (%) | ω (cm$^{-1}$) | k-points | Modes |
|---|---|---|---|
| -4 | 0.3i | Γ | FE$_{xy}$ |
|  | 259i | Γ | FE$_z$ |
|  | 162i | M | AFD$_{xy}$ |
|  | 192i | M | AFD$_{zi}$ |
|  | 207i | M | AFD$_{zo}$ |
| 0 | 101i | Γ | FE$_{xy}$ |
|  | 96i | Γ | FE$_z$ |
|  | 137i | M | AFD$_{xy}$ |
|  | 82i | M | AFD$_{zi}$ |
|  | 119i | M | AFD$_{zo}$ |
| 4 | 230i | Γ | FE$_{xy}$ |
|  | 91 | Γ | FE$_z$ |
|  | 118i/62i$^*$ | M | AFD$_{xy}$ |
|  | 83 | M | AFD$_{zi}$ |
|  | 38i | M | AFD$_{zo}$ |

*: There are more than one unstable modes corresponding to a certain distortion-type.

Table II. Calculated polarizations (in μC/cm$^2$) of FE modes and magnitudes (in degree) of AFD modes for 1:1 *Pc* and *P4bm* BTO/CTO superlattice on [001] STO substrate.

| Modes | FE$_{xy}$ | FE$_z$ | AFD$_{xy}$ | AFD$_{zi}$ | AFD$_{zo}$ |
|---|---|---|---|---|---|
| *Pc* | 15.8 | 24.0 | 3.4 | 2.3 | 3.1 |
| *P4bm* | 0.0 | 19.6 | 0.0 | 0.8 | 6.0 |

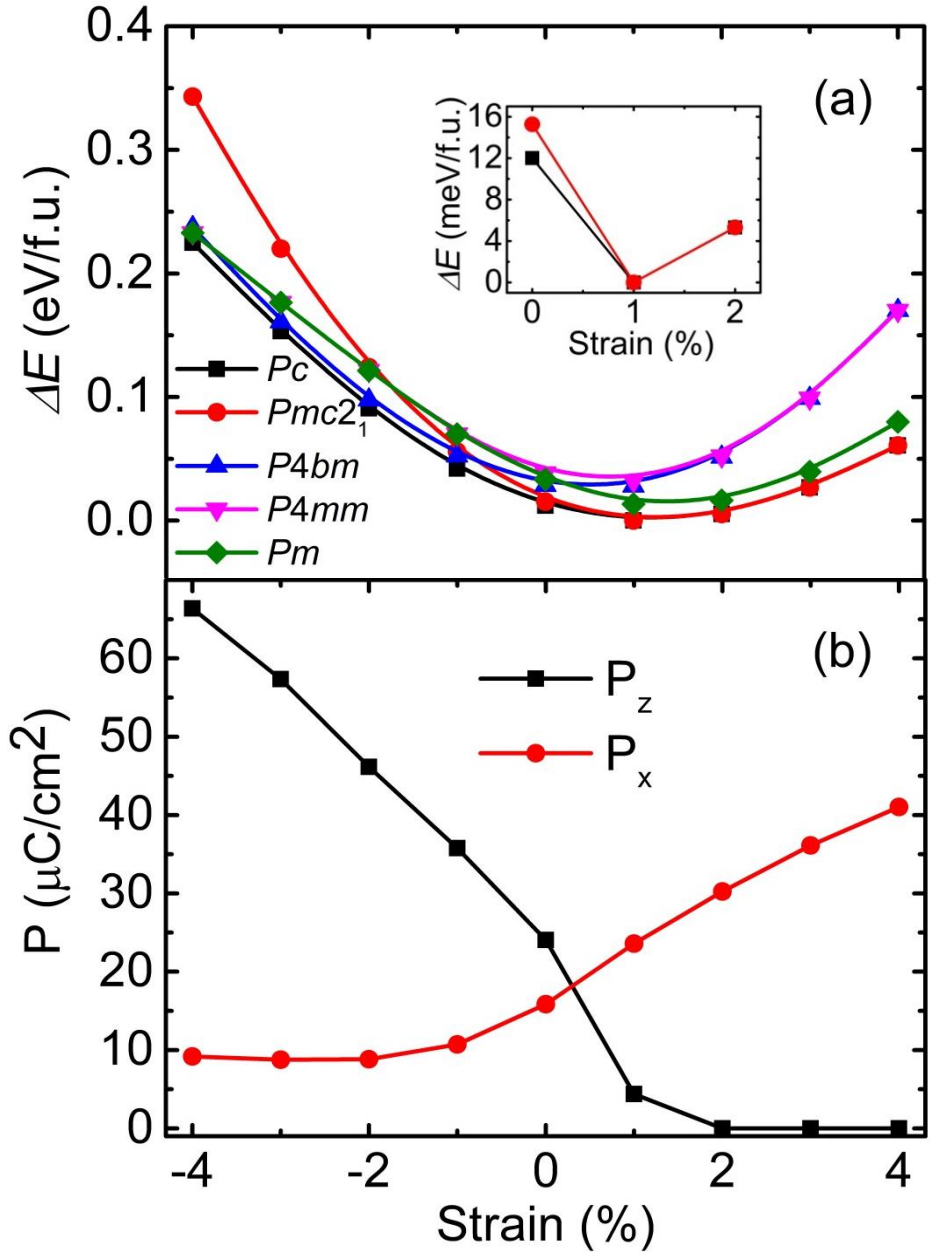

Figure 1. (a) Calculated total energies versus the epitaxial strain (-4% to 4%) for 1:1 BTO/CTO superlattices. Previously suggested *P4bm*, *P4mm* and *Pm* phases are also shown for comparison. Inset shows a zoomed view in the strain range of 0-2%. (b) Calculated ferroelectric polarizations ($P_y = P_x$) of the lowest energy phases at each strain.

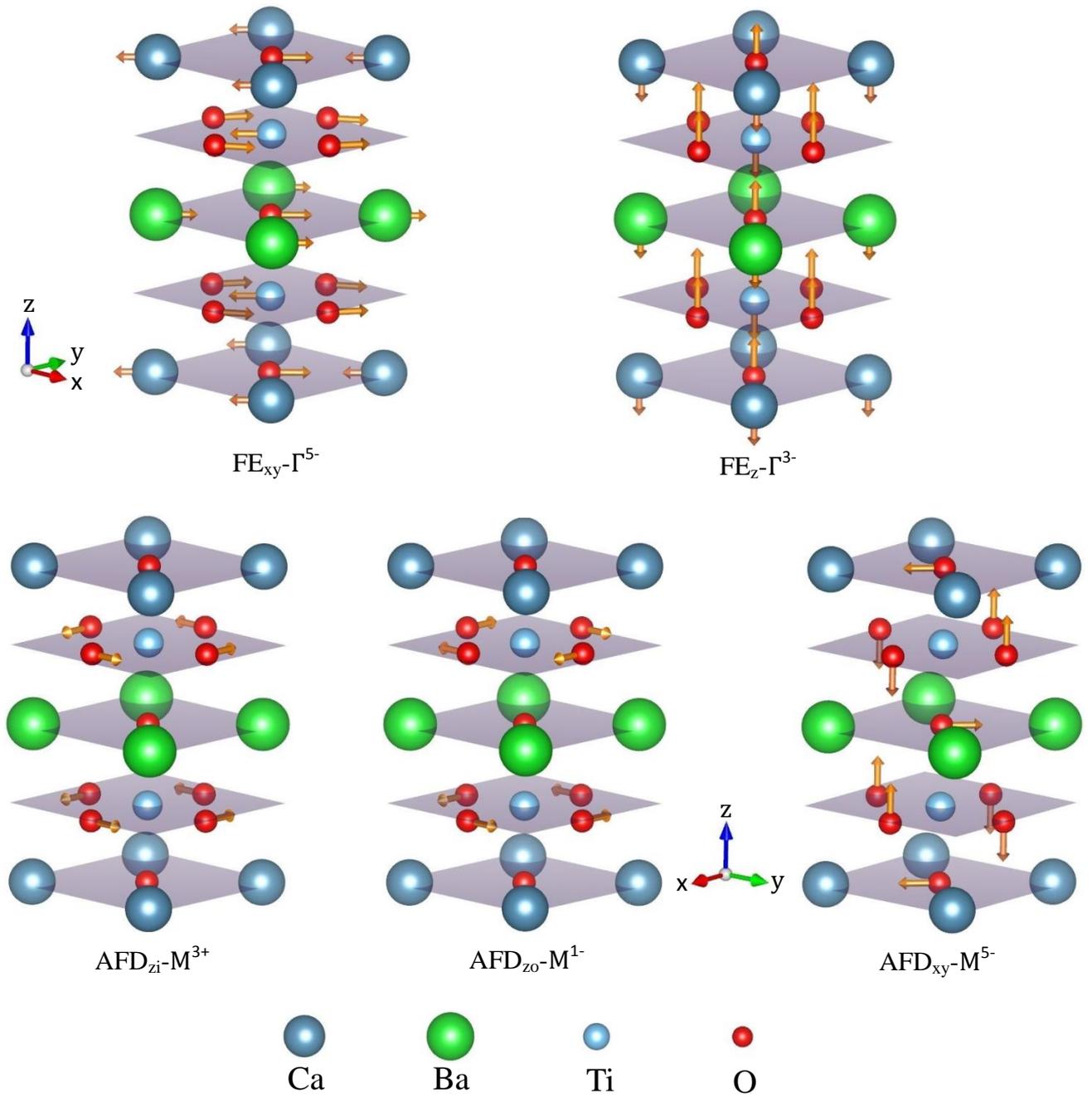

Figure 2. Schematic view of the atomic displacements of unstable modes at Γ and M in *P4mmm* structure of a 1:1 BTO/CTO superlattice on [001] STO substrate.

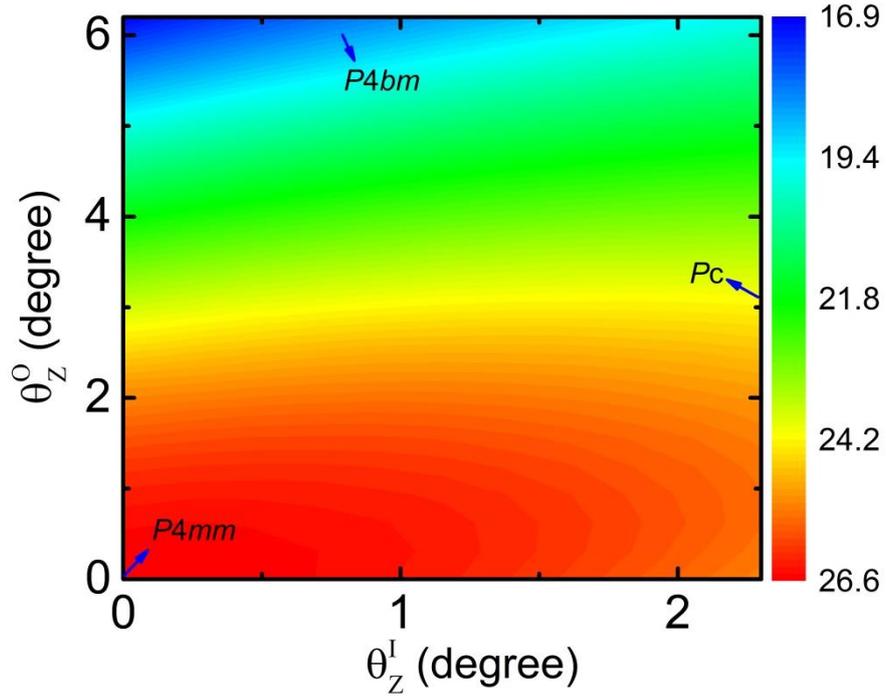

Figure 3. Contour plot of FE$_z$ polarization as a function of $\theta_Z^I$ and $\theta_Z^O$, where $\theta_Z^I$ and $\theta_Z^O$ are defined as the magnitudes of AFD$_{zi}$ and AFD$_{zo}$ rotations, respectively. FE$_z$ polarization (in μC/cm$^2$) gradually increases from blue color (16.9 μC/cm$^2$) to red color (26.6 μC/cm$^2$). *Pc*, *P4bm* and *P4mm* indicate the magnitudes of AFD$_{zi}$ and AFD$_{zo}$ rotations in each case.

(a)

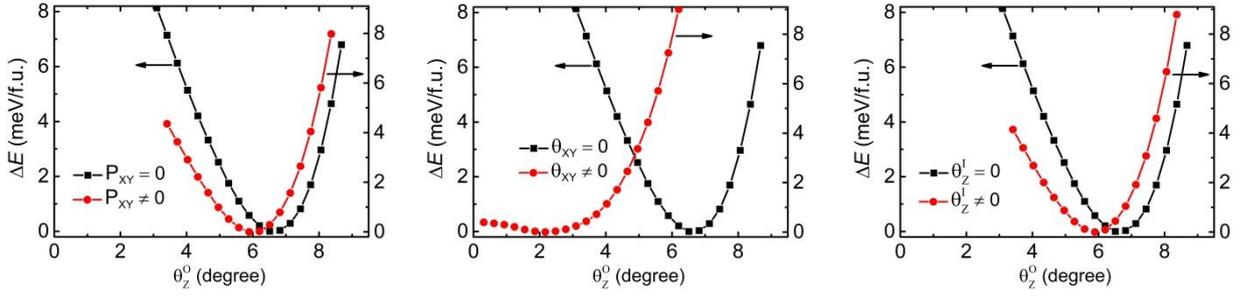

(b)

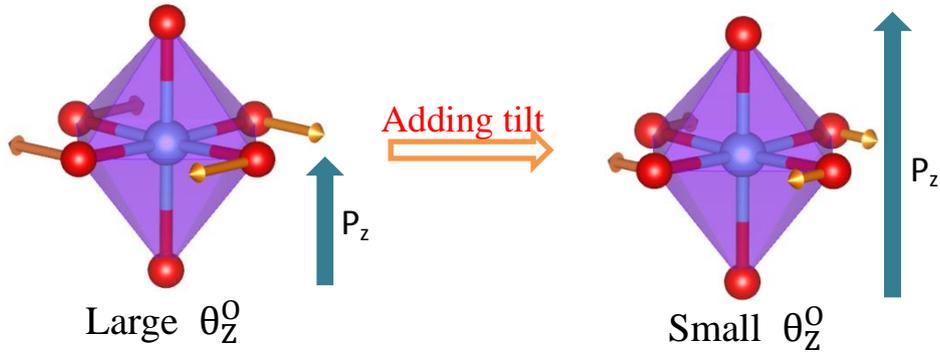

Figure 4. (a) Energy as a function of the rotation of oxygen octahedron with (right axis) and without (left axis) $FE_{xy}$, $AFD_{xy}$ and $AFD_{zi}$ modes. (b) Schematic illustrations of the mechanism of enhancing $FE_z$ polarization through suppressing $AFD_{zo}$ rotations by $AFD_{xy}$ tilts. The yellow arrows indicate the displacements of the oxygen atoms and the blue ones indicate $FE_z$ polarizations.

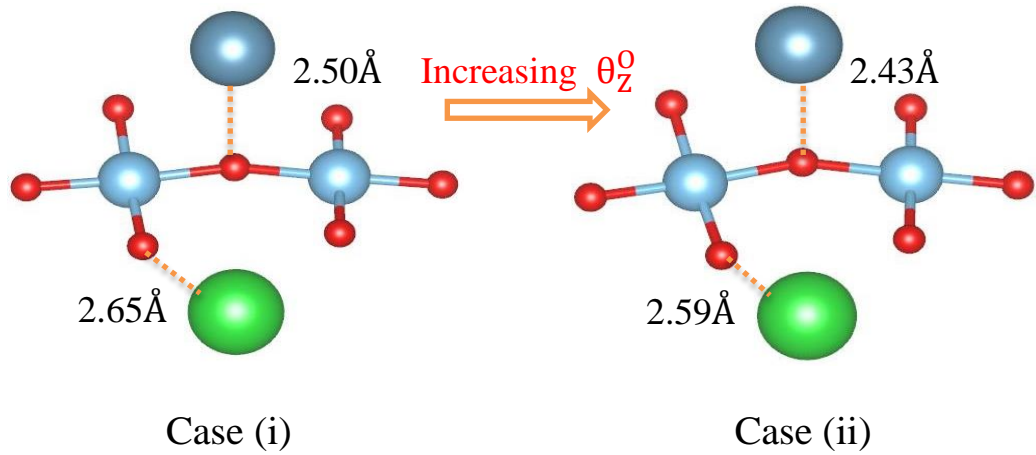

Figure 5. Bond lengths of Ba-O and Ca-O bonds comparison between the two cases (i): $\theta_Z^O = 3.1°$ and $\theta_{XY} = 3.4°$; (ii): $\theta_Z^O = 6.2°$ with $\theta_{XY}$ as in case (i), where $\theta_{XY}$ represent the magnitude of the tilt of oxygen octahedron. Only Ba-O bond shorter than 2.75 Å (summation of the Bi and O ions' ionic radius) and the shortest Ca-O bond in the system are shown.

Supplementary Materials for

# Polarization Enhancement in Perovskite Superlattices by Oxygen Octahedral Tilts

X. Z. Lu[a], X. G. Gong, and H. J. Xiang[b]

Key Laboratory of Computational Physical Sciences (Ministry of Education), State Key Laboratory of Surface Physics, and Department of Physics, Fudan University, Shanghai 200433, P. R. China

e-mail: a) xuezenglu11@fudan.edu.cn; b) hxiang@fudan.edu.cn

**1. Phase diagram and physical properties for 2:2 BaTiO$_3$/CaTiO$_3$ superlattices.**

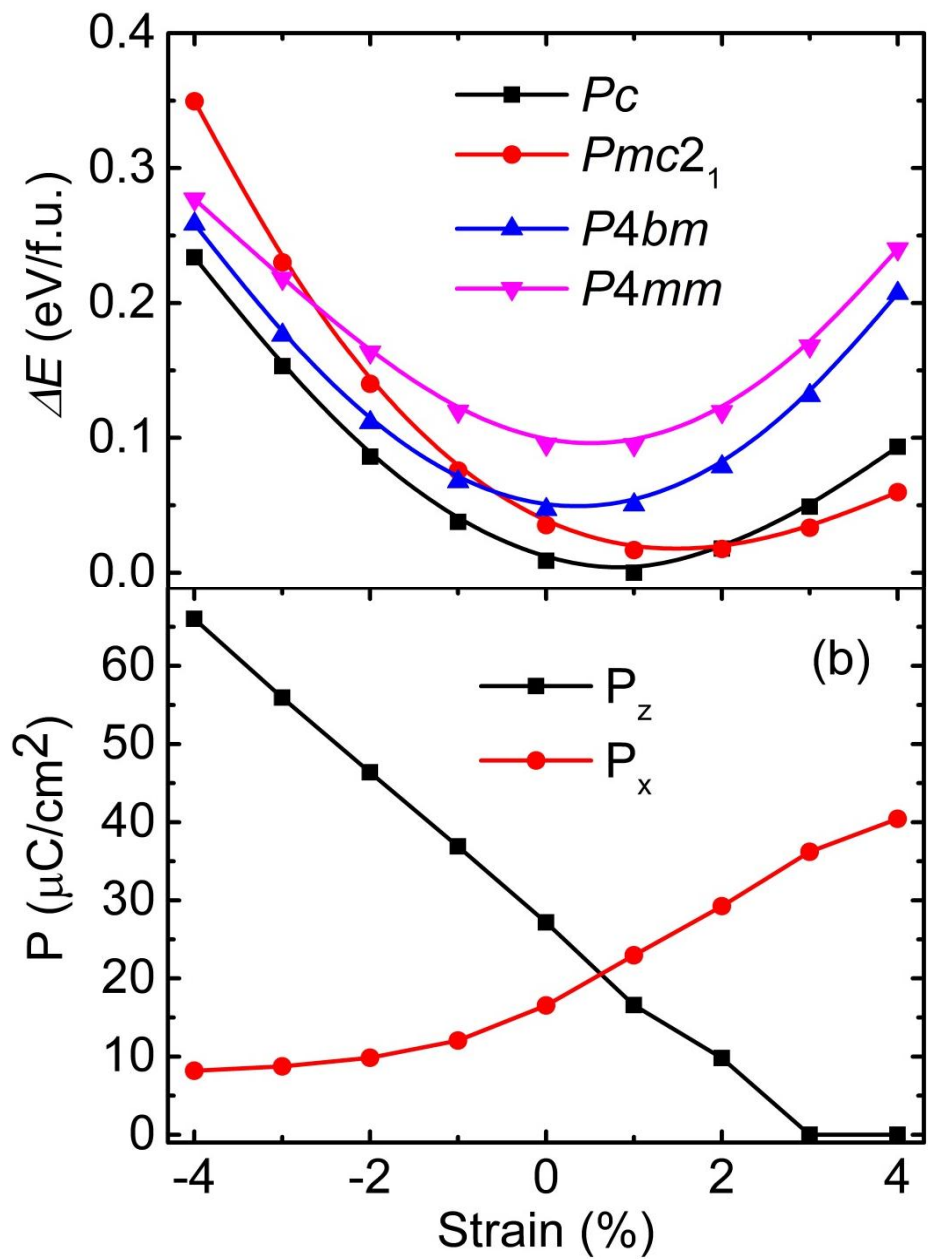

Figure S1. (a) Calculated total energies versus the epitaxial strains in the range of -4-4% for 2:2 BTO/CTO superlattices. Only energy-lowest phases are shown except for *P*4*bm* and *P*4*mm* phases for comparison. (b) Calculated ferroelectric polarizations (P$_y$ = P$_x$) of the lowest energy phases at each strain.

Table SI. Calculated frequencies of modes [1] at Γ and M in *P4mmm* structure of 2:2 BTO/CTO superlattices at different strains.

| Strain (%) | ω (cm$^{-1}$) | k-points | Modes |
|---|---|---|---|
| -4 | 110i | Γ | FE$_{xy}$ |
| | 268i | Γ | FE$_z$ |
| | 204i | M | AFD$_{xy}$ |
| | 270i/198i/76i[*] | M | AFD$_{zi}$ |
| | 197i | M | AFD$_{zo}$ |
| 0 | 140i | Γ | FE$_{xy}$ |
| | 131i | Γ | FE$_z$ |
| | 181i | M | AFD$_{xy}$ |
| | 210i/100i[*] | M | AFD$_{zi}$ |
| | 98i | M | AFD$_{zo}$ |
| 4 | 258i/185i/121i/43i[*] | Γ | FE$_{xy}$ |
| | 38 | Γ | FE$_z$ |
| | 158i/67i[*] | M | AFD$_{xy}$ |
| | 172i | M | AFD$_{zi}$ |
| | 65 | M | AFD$_{zo}$ |

*: There are more than one unstable modes corresponding to a certain distortion-type.

In this 2:2 superlattices, the same space groups as those in 1:1 superlattices for the ground states can be found, i.e., *Pc* phase in the strain range of -4-2% and *Pmc*2$_1$ phase with the strain higher than 2%. *Pc* phase is lower in energy by average 36 meV per formula unit with respect to *P4bm* phase. Concerning the origin of the polarization in 2:2 superlattices, the same conclusion can be drawn for FE$_z$ component from Fig. S1b and Table SI as that in 1:1 superlattices, that is, FE$_z$ polarization is proper ferroelectricity. For $FE_{xy}$ polarization, different from the evolution of the polarization with the strain in 1:1 superlattices, it is also most likely to result from the polarization-strain coupling like $FE_z$ polarization. Therefore in the whole strain range studied, $FE_{xy}$ polarization is proper ferroelectricity in 2:2 superlattices due to the mode itself is significantly unstable.

Table SII. Calculated FE polarizations (in $\mu C/cm^2$) for 2:2 *Pc* and *P4bm* BTO/CTO superlattices on [001] STO substrate.

| Modes | $FE_{xy}$ | $FE_z$ |
|-------|-----------|--------|
| *Pc*    | 16.5      | 27.2   |
| *P4bm*  | 0.0       | 19.1   |

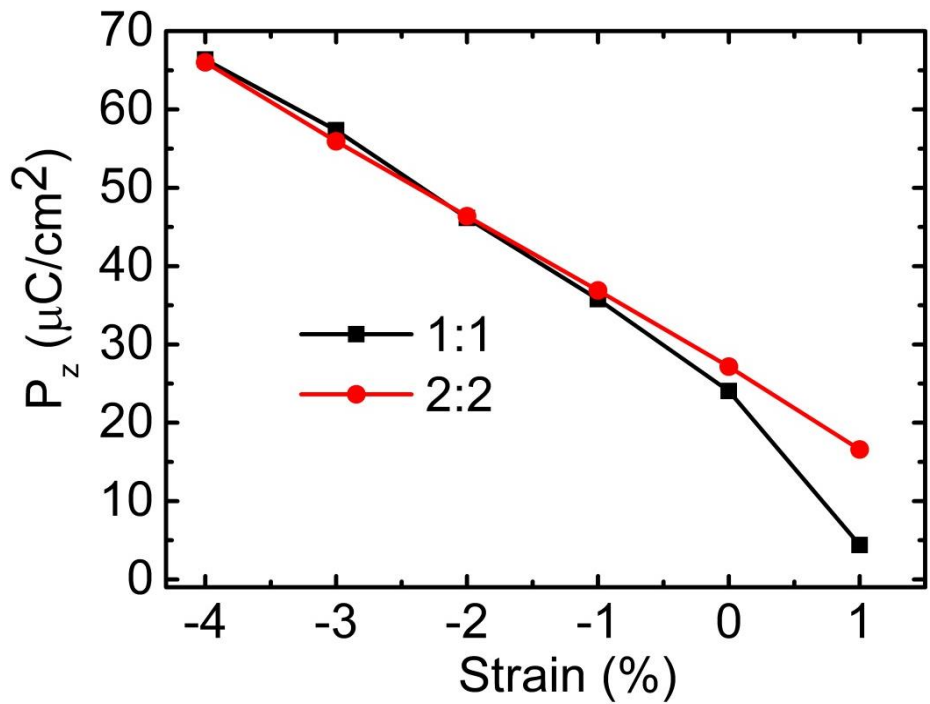

Figure S2. Comparison between 1:1 and 2:2 $FE_z$ polarization in the strain range of -4 – 1%.

From Table SII, we can find that $FE_z$ polarization in 2:2 *Pc* superlattice is much larger than that in *P4bm* superlattice and even slightly larger than that in 1:1 *Pc* superlattice, which is different from the experiments [2,3] where $FE_z$ polarizations of 1:1 and 2:2 superlattices have a similar value of ~8.5μC/cm$^2$ when grown on STO substrate. As can been seen in Fig. S2, the difference between $FE_z$ polarizations of 1:1 and 2:2 superlattices becomes much smaller when the compressive strain larger than 1%, at which the results show qualitatively agreement with the experiments [2,3]. But the values around 36μC/cm$^2$ at -1% strain are much larger than the experiments [2,3]. The quantitative discrepancy may be due to the difference between experimental and theoretical conditions such as the effects of strain and temperature. In our study, we performed DFT calculations at 0K and use the theoretical lattice constant of bulk STO (3.856 Å), compared with the room temperature and experimental value of bulk STO lattice constant (3.905 Å). Besides, using the short-time pulse measurement [2], $FE_z$ polarization of 1:1 superlattices is observed to be ~14μC/cm$^2$. Thus, the experimental observation of our results remains to be confirmed.

## 2. The effects of FE$_{xy}$ and AFD$_{xy}$ modes on FE$_z$ mode.

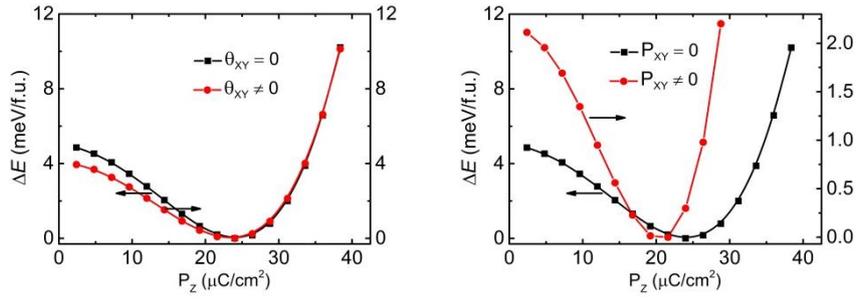

Figure S3. Energy as a function of the out-of-plane polarization with (right axis) and without (left axis) FE$_{xy}$ and AFD$_{xy}$ modes, respectively.

## 3. Coupling between $AFD_{zo}$ and $AFD_{xy}$ modes.

Table SIII. Values (in meV) of the coefficients in the model described below.

| α | β | γ |
|---|---|---|
| -2.58 | 2.92×10$^{-2}$ | 0.20 |

The free energy can be expanded in terms of the order parameters of $FE_{xy}$, $FE_z$, $AFD_{xy}$, $AFD_{zi}$ and $AFD_{zo}$ modes in the Landau expansion. To describe the coupling between $AFD_{zo}$ and $AFD_{xy}$ modes, we only need to consider the following terms:

$$E = \alpha(\theta_Z^O)^2 + \beta(\theta_Z^O)^4 + \gamma(\theta_Z^O \theta_{XY})^2 + \text{the others}$$

where α, β and γ are coefficients; $\theta_Z^O$ and $\theta_{XY}$, in degree, label the magnitudes of $AFD_{zo}$ rotation and $AFD_{xy}$ tilt, respectively. After the numerical DFT calculations, the values of the coefficients are obtained, as listed in Table SIII. For the coupling between $AFD_{zo}$ and $AFD_{xy}$ modes, $\alpha + \gamma\theta_{XY}^2 < 0$ when $\theta_{XY}=3.4°$, which indicates that $AFD_{zo}$ rotation will be not completely suppressed by $AFD_{xy}$ tilt but with a value of 2.1° when $\theta_{XY}=3.4°$, consistent with the direct DFT calculations (see Fig. 4a of the text).

**References**

1. At the three strains, there are more than the five unstable modes at Γ and M of *P4mmm* structure, we only show the five modes for three reasons: i), they are more likely to be condensed according to our analysis; ii), to show the evolution of the frequency of the modes with the strain; iii), they are related to the ferroelectric distortions and oxygen octahedral rotations and tilts.